\documentclass[showpacs,preprintnumbers,amsmath,amssymb,twocolumn]{revtex4} 
\usepackage{graphicx}
\usepackage{makeidx}
\usepackage{bm}

\def\epl#1#2#3{{ Europhys. Lett.} {\bf #1}, #2 (#3)}
\def\prl#1#2#3{{ Phys. Rev. Lett.} {\bf #1}, #2 (#3)}

\def\pla#1#2#3{Phys. Lett. A {\bf #1}, #2 (#3)}
\def\ijbc#1#2#3{Int J. Bifurcation and Chaos {\bf #1}, #2 (#3)}
\def\pre#1#2#3{Phys. Rev. E {\bf #1}, #2 (#3)}

\def\pr#1#2#3{Phys. Rep. {\bf #1}, #2 (#3)}

\def\pnas#1#2#3{Proc. Natl. Acad. Sci. (USA) {\bf #1}, #2 (#3)}

\def\pr#1#2#3{Phys. Rev. {\bf #1}, #2 (#3)}
\def\jpa#1#2#3{J. Phys. A {\bf #1}, #2 (#3)}

\def\jas#1#2#3{J. Atmos. Sci. {\bf #1}, #2 (#3)}

\def\jns#1#2#3{J. Nonlin. Sci. {\bf #1}, #2 (#3)}
\def\cmp#1#2#3{Comm. Math. Phys. {\bf #1}, #2 (#3)}
\def\physd#1#2#3{Physica D {\bf #1}, #2 (#3)}
\def\physa#1#2#3{Physica A {\bf #1}, #2 (#3)}

\def\ch#1#2#3{Chaos {\bf #1}, #2 (#3)}
\def\nat#1#2#3{Nature {\bf #1}, #2 (#3)}
\def\nl#1#2#3{Nonlinearity {\bf #1}, #2 (#3)}

\def\etal{{\it et al.}~}
\def\epsilon{\varepsilon}

\def\half{\frac{1}{2}}

\def\beqr{\begin{eqnarray}}
\def\eqnr{\end{eqnarray}}
\def\beq{\begin{equation}}
\def\bc{\begin{center}}
\def\ec{\end{center}}
\def\eqn{\end{equation}}
\begin{document}
\title{Design strategies for the creation of aperiodic nonchaotic attractors}
\author{Amitabha Nandi$^1$, Sourav K. Bhowmick$^2$, Syamal K. Dana$^2$ and 
Ram Ramaswamy$^3$}
\affiliation{$^1$Max-Planck Institut f\"ur Physik Komplexer Systeme,
N\"othnitzer Str. 38 01187 Dresden, Germany \\
$^2$ Instrument Division, Indian Institute of Chemical Biology, Kolkata 
700 032, India \\
$^3$School of Physical Sciences, Jawaharlal Nehru University,
New Delhi 110 067, India}
\begin{abstract}
Parametric modulation in nonlinear dynamical systems can give rise to 
attractors on which the dynamics is aperiodic and nonchaotic, namely with 
largest Lyapunov exponent being nonpositive. We describe a procedure for 
creating such attractors by using random modulation or pseudo-random binary 
sequences with arbitrarily long recurrence times. As a consequence the 
attractors are geometrically fractal and the motion is aperiodic on 
experimentally accessible timescales. A practical realization of such 
attractors is demonstrated in an experiment using electronic circuits.
\end{abstract}
\maketitle

\section*{}
{\bf Generating dynamics which is aperiodic, but nevertheless stable in a sense that nearby trajectories
coalesce and synchronize, has been of considerable interest in the past few years. Such motion has been
typically observed in driven dynamical system, in particular when the drive is quasiperiodic.
Since quasiperiodicity is difficult to achieve in practice, a major issue in this regard has been
whether such dynamics can  be achieved by other techniques, and has been answered with various degrees of
success. Here we present a simple design scheme that uses pseudo-random binary sequences with very
long recurrence times to switch the dynamics between two different states. The resultant dynamics goes to an
attractor which is aperiodic and stable, namely has negative Lyapunov exponent. Characterization of such
dynamics reveals the fractal nature of such dynamics and also their differences with the ones obtained by
quasiperiodic drive. Such a design scheme is further realized in an experimental setup using
electronic circuits, suggesting potential applications in practical situations.}

\section{Introduction}
The design of dynamical systems in which the motion is both aperiodic and 
stable has been an objective of a number of recent studies \cite{pr1}. One 
realization of this goal is in strange nonchaotic attractors (SNAs) that 
can be created in quasi--periodically driven systems \cite{prasad,pr1}. In 
such systems which have been known for over twenty years now \cite{gopy}, the 
attractor has a fractal geometry, and this results in the dynamics being 
aperiodic. The Lyapunov exponents of the drive system are nonpositive: this 
lead to a lack of sensitivity to initial conditions, and thus to the 
synchronization of orbits \cite{rr}.

Achieving quasiperiodicity is simple in principle, but difficult in practice. 
Quasiperiodic modulation requires that the system be driven either through a 
single source that  has an irrational frequency (for maps), or with two 
sources whose frequencies are incommensurate (for flows). Experimental 
uncertainties are usually larger than the precision with which rational or 
irrational numbers can be measured, and therefore finding or creating SNAs in 
practical situations has proved to be difficult, except for isolated 
experiments \cite{ditto}. If strange nonchaotic dynamics is to be taken 
seriously, it is necessary to ask if such behaviour can arise in the absence 
of strict quasiperiodicity. In particular, most physical, biological and 
engineering systems will not be quasiperiodic, so it is natural to ask 
whether SNAs can arise in situations where the underlying system is 
autonomous, or has other time dependence. 

Earlier studies have addressed this issue \cite{rajasekhar,sosnov,feudel,lai} 
with varying degrees of success. The use of an external noise source typically 
has the effect of smearing out attractor structure and gives effective 
Lyapunov exponents that are nonnegative. As a result the dynamics is neither 
truly strange nor truly nonchaotic \cite{apinsa}. The recent suggestion by 
Wang \etal \cite{lai}, that additive noise alone can be used to induce robust 
SNAs in both maps and flows  appears to ensure that the attractors so 
created share the mathematical properties of SNAs formed by other bifurcation 
routes \cite{pr1,lai}. 

The approach taken in the present paper pursues a different route. We show 
that by using parametric modulation based on deterministic pseudorandom 
dynamics, it is possible to create dynamical attractors that are  stable, 
nonchaotic and aperiodic, and which will appear strange on any  measurable 
timescale. In this context the term `strangeness'  refers purely to the 
geometry, while stability refers to the  fidelity in signal reproduction or 
tracking \cite{raj}, an issue that, for instance, underlies schemes for 
encryption and communication. 

Our procedure for the creation of such aperiodic nonchaotic attractors (ANAs) 
relies on the use of binary random numbers to modulate the dynamical 
system between two dynamical states, a stable fixed point and a 
chaotic attractor. By suitably designing the drive dynamics, it is possible 
to ensure that the asymptotic dynamics of the driven system has (a) a 
negative largest Lyapunov exponent, and (b) nontrivial and complicated 
geometry on spatial scales that are determined by the (essentially) 
experimental resolution. We show that such attractors can be created in both 
discrete-time autonomous maps as well as flows, and further present an 
experiment based on electronic circuits to support our findings. This suggests 
that ANAs could have potential application in practical situations where 
aperiodic dynamics is desirable, as for example in chaotic communications. At 
the same time,  these attractors have differences from those created by 
quasiperiodic driving and other methods \cite{pr1,prasad}. It should be 
noted that dichotomous driving has been used before in both numerical 
\cite{raja} as well as experimental \cite{brousell} studies, although 
the motivation there was to study noise--induced transitions between 
different states or attractors.

In the following section we discuss the design of ANAs in the driven H\'enon 
map and the driven Lorenz system  using a deterministic feedback shift 
register to generate a pseudorandom drive \cite{pr1}. In Sec~\ref{sec-cs} we 
discuss similar driving mechanism using chaotic sequence from a Chua circuit. 
The study of such attractors and their proper characterization is discussed in 
Sec~\ref{sec-char}, where we show the geometric differences between ANAs and 
comparable SNAs. In Sec~\ref{sec-exp} we present an experimental realization 
of such attractors using electronic circuits, and  conclude  in 
Sec~\ref{sec-dis} with a discussion and summary of our results.

\section{Dichotomous deterministic modulation}
\label{sec2}
Consider a dynamical system (with one freedom and a single parameter for 
simplicity, but with obvious extension to higher dimensions and to the case 
of several parameters)
\beq
\label{drive-scheme}
x  \to f(x, b),
\eqn
that is modulated through the output of a binary drive sequence (strings of 
0's and 1's) 
$z_n$ as
\beq
\label{xeq}
x_{n+1} =  f(x_{n},b_1+z_{n}(b_2-b_1)).
\eqn
Depending on the value of $z_{n}$, the system parameters thus switch 
between $b_1$ and $b_2$, giving a dichotomous modulation that is, nevertheless,
deterministic. 

One standard way of achieving this is to use a linear feedback shift register 
(LFSR)\cite{simon} that generates a pseudo-random bit sequence $\{z\}$ 
through a delay mapping of the general form 
\beq
\label{lfsr}
z_{n+1} = \sum_{i=1}^N  a_{i}z_{n+1-i}~~~~~\mbox{mod~2},
\eqn
where $a$ is also a binary variable. For a specific choice of nonzero $a$'s 
for  a given $N$ (the ``tap sequence"), the dynamics is on an attractor with  
period $ \le 2^N$-1. The analog generalization (namely, the analog feedback 
shift registers (AFSR)) \cite{gg} uses the same coefficients in a  
continuous mapping
\beq
\label{afsr}
z_{n+1} = \half - \half \cos \pi \sum_{i=1}^N a_{i}z_{n+1-i}, 
\eqn
to generate a pseudorandom sequence of 0's and 1's.  This dichotomous drive 
is, importantly, a dynamical system and the drive sequence is an 
{\it attractor} of the dynamics. The sequence is optimal if arbitrarily long 
sequences  of either 0 or 1 occur. The theory of LFSRs (and thus of AFSRs) is 
well-developed and minimal tap sequences that produce the longest possible 
(namely 2$^N$-1) period pseudorandom sequences are easily available 
\cite{simon}. For sufficiently large $N$, the period of the pseudorandom 
sequence can quickly exceed the age of the universe at any realistic sampling 
rate.

Designing aperiodic but nonchaotic dynamics in $x$ is straightforward: for 
instance if $b_1$ corresponds to a case of, say, superstable dynamics, 
and $b_2$ to the case of chaotic dynamics in the system (Eq.~\ref{xeq}), the 
resultant dynamics in the driven system will be aperiodic but will rapidly be 
attracted to the superstable orbit whenever there is a ``gap'', namely a 
string of 0's. The Lyapunov exponent will consequently be negative and as a 
result trajectories with arbitrary initial conditions will synchronize.

Special attention should be given when choosing the tap sequence $N$. It 
must be large enough such that the recurrence time, namely $2^N-1$ is much 
longer than the time-scales used for simulations. For smaller $N$, the recurrence 
also become short, and since the AFSR dynamics is periodic, the system 
dynamics will go to a periodic attractor. 

Any other random sequence will also serve the purpose, but LFSRs or AFSRs offer a 
practical advantage over other pseudo random number generators (PRNGs). 
The shift registers are maximally stable \cite{gg}: being attractors of the 
dynamics their stability and controllability---unlike that of stochastic 
sequences or  PRNGs---is more easily ensured. Furthermore, since feedback 
shift registers are dynamical systems as well,  the entire drive--response 
system can be represented as a delay dynamical system.

We discuss representative examples below.  

\subsection{H\'enon Map}
\label{henon}
\begin{figure}
\scalebox{0.35}{\includegraphics{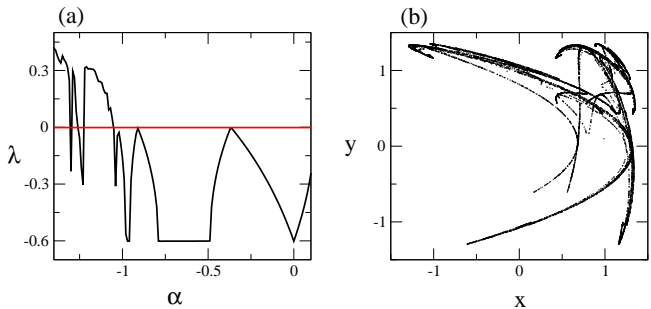}}
\caption{
(a) Variation of the Lyapunov exponent with $\alpha$ at $\beta=0.3$ for the 
unmodulated H\'enon map. (b) Aperiodic nonchaotic orbit in the H\'enon map, 
parametrically modulated through an AFSR of order $N$=24. The attractors are 
obtained for parameter switching between $\alpha=-0.14$ and  $\alpha=-1.2$. 
The largest Lyapunov exponent is $\lambda_1=-0.16$. The recurrence time for 
the AFSR is $2^{24} - 1$.}
\label{fig1}
\end{figure}

The H\'enon map \cite{henon} is a well studied two dimensional iterative
dynamical system, given by the following equations
\beqr
\label{hen1}
x_{n+1} &=& 1 + \alpha x_{n}^{2} + y_{n}, \\
y_{n+1} &=& \beta x_{n}.
\label{hen2}
\eqnr
For certain choice of $(\alpha,\beta)$ and for a certain range of initial 
condition, the system exhibits chaotic or fixed point or periodic behaviour 
(see Fig.~\ref{fig1}a).
 
Applying the strategy discussed above produces attractors which are both 
nonchaotic and essentially strange. In the H\'enon map, at $\alpha=-1.2$ the 
dynamics is chaotic and at $\alpha=-0.14$ the dynamics goes to a fixed point 
(see Fig.~\ref{fig1}a). Combining Eq.~(\ref{afsr}) and 
Eqs.~(\ref{hen1}-\ref{hen2}), we get the following equations for the 
drive--response system
\begin{eqnarray}
\label{henafsr1}
x_{n+1} &=& 1 + \alpha(1 + cz_{n})x_{n}^2 + y_n, \\
y_{n+1} &=& \beta x_n, \\
z_{n+1} &=& \half[1 - \cos(\pi \sum_{i=1}^N a_{i}z_{n+1-i})].
\label{henafsr2}
\end{eqnarray}
If we choose $\alpha=-0.14$, $\beta=0.3$ and $c=\frac{53}{7}$, then depending 
on whether $z_{n}$ is $0$ or $1$, the quantity $\alpha(1 + cz_{n})$ will take 
values of either $-0.14$ or $-1.2$. As a result the dynamics hops between two 
different states such that the global dynamics is stable and nonchaotic. The 
Lyapunov exponent for the global dynamics is given by $\lambda=-0.16$. 
Fig.~\ref{fig1}(b) shows the attractor in phase space. Clearly the attractor 
looks geometrically strange; the dynamics is aperiodic and nonchaotic. 

\subsection{Lorenz system}
\label{l-s}
The same strategy can be applied to a flow: switching the dynamics between a 
fixed-point or limit--cycle and a chaotic attractor can result in such strange 
dynamics in a modulated Lorenz system \cite{Lorenz},
\begin{eqnarray}
\label{lorenz1}
\dot{x} &=& \sigma(y-x),\\
\dot{y} &=& (1+c\zeta(t)) \rho x -y-xz,\\
\dot{z} &=& xy-\beta z. 
\label{lorenz2}
\end{eqnarray}
which has the parameter $\rho$ changing in a time--dependent manner through 
the variable $\zeta$,
\beq
\label{zeta}
\zeta(t)=z_n~~~~~~n\tau \ge t \ge (n-1) \tau.
\eqn
As in the mapping in Sec~\ref{henon} above, $z_n$ is the output of an 
AFSR (Eq.~\ref{afsr}) and the switch duration $\tau$ is an additional 
parameter in the problem: it is the time for which a trajectory is switched 
into either of the states. In the present problem we have taken $\tau$ 
to switch into either of the states to be equal, but one can choose 
mismatched $\tau$'s.

\begin{figure}
\scalebox{0.22}{\includegraphics{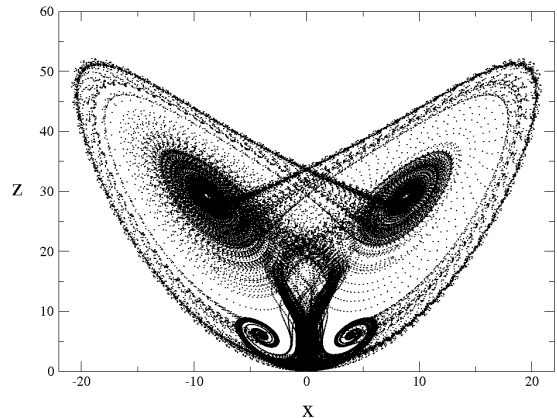}}
\caption{
Aperiodic nonchaotic attractor obtained in the phase-space for the Lorenz 
system under the AFSR modulation. Dynamics obtained for parameter values 
$\rho=7$ and $c=\frac{23}{7}$ such that $\rho$ switches between 
the values 7 and 30. Switch duration time is chosen to be 
$\tau=10$, and the largest Lyapunov exponent is $\lambda_1=-0.00985$.}
\label{fig2}
\end{figure}
The attractor shown in Fig.~\ref{fig2} is the result of the dynamics hopping 
between parameter values $\rho=7$ and $\rho=30$. At the latter value the 
attractor is chaotic with the characteristic butterfly structure about two 
symmetric unstable fixed points $\lbrace{-8.79,-8.79,29}\rbrace$ and 
$\lbrace{8.79,8.79,29}\rbrace$. For $\rho=7$ the system has two symmetrical 
attractive fixed points at $\lbrace{-4,-4,6}\rbrace$ and 
$\lbrace{4,4,6}\rbrace$. When the parameter switches between the values, the 
dynamics alternates between the stable and the unstable fixed point dynamics, 
resulting in the structure visible in Fig.~\ref{fig2}. 

\section{Chaotic modulation}
\label{sec-cs}
A sequence generated from a chaotic signal can also generate aperiodic 
nonchaotic dynamics. For instance, consider the Chua oscillator \cite{chua} 
which is described by the following sets of equations
\begin{eqnarray}
\label{chua1}
\dot{x} &=& c_1(y-x-g(x)),\\
\dot{y} &=& x-y+z,\\
\dot{z} &=& -c_2y-c_3z,
\label{chua2} 
\end{eqnarray}
where $g(x)$ is given by, 
\begin{eqnarray}
\label{chua11}
&&m_1x+m_1-m_0~~~~~\mbox{if }x\le-1,\\
g(x)=&&m_0x~~~~~~~~~~~~~~~~~~~~\mbox{if }-1\le x\le1,\\
&&m_1x+m_0-m_1~~~~~\mbox{if }1\le x. 
\label{chua22}
\end{eqnarray}

\begin{figure}
\scalebox{0.32}{\includegraphics{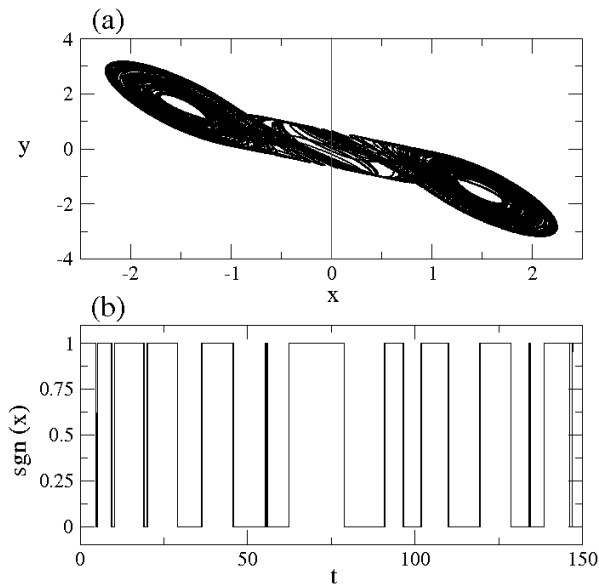}}
\caption{
(a)The Chua double-scroll attractor in the $x-z$ plane. The fixed 
parameter values are $c_1=9$, $c_3=0$, $m_0=-\frac{8}{7}$, $m_1=\frac{5}{7}$. 
The double-scroll is obtained by tuning the parameter $c_2$. Here we take 
$c_2=14.141$. (b) Corresponding bit sequence obtained assigning all points 
with positive value of $x$ as `1' and with negative values as `0'.}
\label{fig3}
\end{figure}

For certain choice of the parameter $c_3$, a chaotic ``double scroll'' 
attractor is obtained (see Fig.~\ref{fig3}a). We extract an indicator 
sequence of binary numbers from this chaotic attractor (see Fig.~\ref{fig3}b) 
using the prescription that whenever the trajectory is on the left 
(resp. right) scroll, the indicator sequence is taken as  1 (resp. 0). 

\begin{figure}
\scalebox{0.22}{\includegraphics{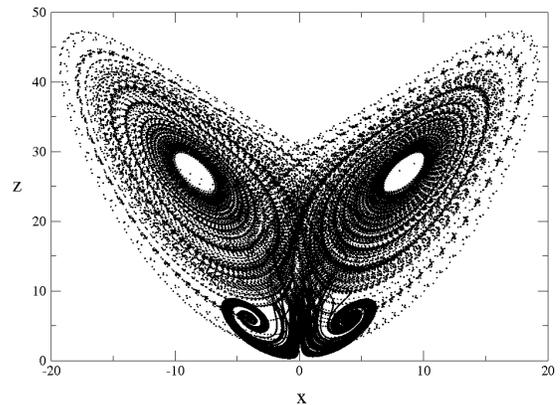}}
\caption{
Aperiodic nonchaotic attractors obtained in the phase-space for the Lorenz 
system using chaotic bit sequence from the Chua double scroll. The 
parameter values are $\rho=7$ and $c=\frac{21}{7}$ such that $\rho$ switches 
between the values 7 and 28. Switch duration time $\tau$ is chosen to be 
unity. The largest LE is $\lambda_1=-0.03$. The other parameters $\sigma$ 
and $\beta$ have values 10 and $\frac{8}{3}$ respectively.}
\label{fig4}
\end{figure}

This gives a drive signal which is piecewise constant, which upon application 
to the Lorenz system discussed in Sec~\ref{l-s}  yields an attractor 
(see Fig.~\ref{fig4}) that is very similar  to that obtained via deterministic 
dichotomous driving. The main feature that both these drive signals share is 
that they have long periods when the drive is on the stable attractor, and 
this suffices to ensure that the  eventual  dynamics is aperiodic, and that 
the attractor has a complicated geometry and a nonpositive largest Lyapunov 
exponent. 

In the following section, we characterize these attractors via measures 
used in the study of SNAs.

\section{Characterization of ANAs}
\label{sec-char}
Computation of the largest Lyapunov exponent shows that these attractors are 
nonchaotic. However, since the attractors are geometrically strange only over 
finite resolution, they  differ in the qualitative and quantitative aspects 
of their local fluctuation properties \cite{pr-fle} from other similar 
attractors such as SNAs. 

\begin{figure}
\scalebox{0.5}{\includegraphics{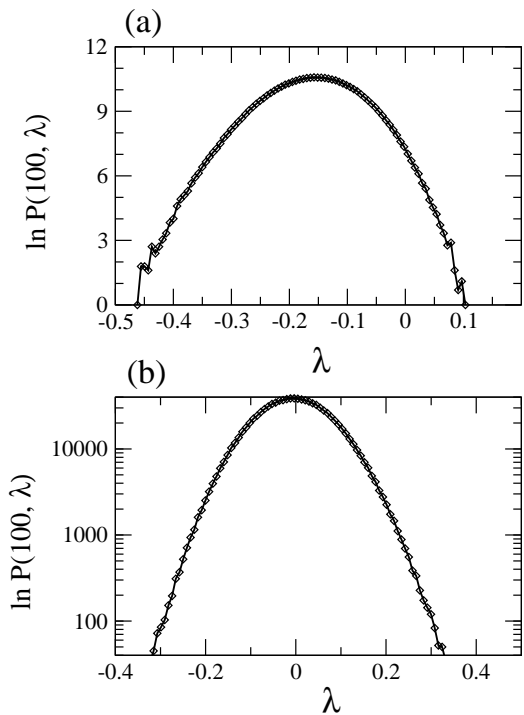}}
\caption{Finite time Lyapunov exponent (FTLE)s. (a) for the modulated H\'enon 
map with parameter values $\alpha=-0.14$ and $c=\frac{53}{7}$. Asymptotic 
largest LE value is $\lambda_1=-0.16$. The distribution is taken for $N=100$. 
(b) Similar results for the modulated Lorenz system with parameter values 
as in Fig~\ref{fig2} and asymptotic largest LE is $\lambda_1=-0.00985$.}
\label{fig5}
\end{figure}
                                                                                
Finite--time Lyapunov exponents (FTLES) \cite{lle} are local estimates for 
the rate of divergence between nearby trajectories, and explicitly depend 
both on the time interval $\tau$ over which they are measured as well as the 
initial conditions. By computing $\lambda_{\tau}$ for a large number of 
initial points in the phase space, one can obtain the stationary distribution,
\beqr
P(\lambda,\tau) &=& \mbox{Probability that }\lambda_{\tau}\\
&&\mbox{lies in the interval~} (\lambda,\lambda+d\lambda).
\eqnr
For SNAs this is typically broad and non-Gaussian \cite{num8,rich2,pr-fle}, 
although the mean of the distribution, namely the asymptotic Lyapunov 
exponent, is negative.

The FTLE distribution for ANAs is purely Gaussian: see 
Figs.~\ref{fig5}((a) and (b))  for the driven H\'enon and Lorenz systems. 
This follows from the pseudo-random nature of the driving which switches the 
dynamics between a regular and a chaotic state. Correlations die out rapidly 
and the FTLEs satisfy the central limit theorem. As a result the distribution 
is a Gaussian whose spread is a function of the length of the trajectory 
\cite{pr-fle}.
\begin{figure}
\scalebox{0.35}{\includegraphics{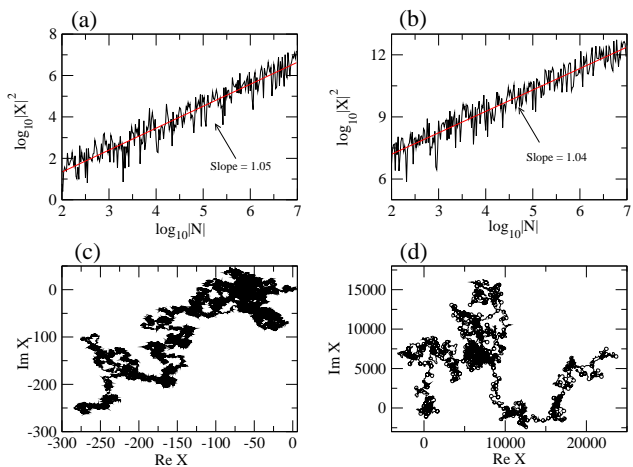}}
\caption{
Finite--time Fourier power $|X(\Omega,N)|^{2}$ versus $T$ on a logarithmic 
scale for (a) the H\'enon map, and (b) the Lorenz system. Fractal walk of the 
spectral  trajectory in the complex plane (ReX, ImX) for (c) the H\'enon map, 
and (d) for the Lorenz system. Parameter values are as in Fig.~\ref{fig1}(a) 
and Fig.~\ref{fig2}.}
\label{fig6}
\end{figure}

The contrast between ANAs and SNAs is also evident in the Fourier spectrum 
\cite{qqq}. Upon computing the time-dependent partial Fourier sum 
\cite{prasad,qqq}
\beq
\label{spectra}
X(\Omega,N) = \sum_{n=1}^N x_{n}\exp^{i2\pi n\Omega},
\eqn
at irrational frequency $\Omega$ the golden mean ratio, $(\sqrt{5}-1)/2$, 
the scaling relation $|X(\Omega,N)|^{2} \sim N^\beta$ is observed. For 
noisy motion, $\beta=1$, and the spectrum is continuous. For periodic motion 
$\beta=2$ and the spectrum is discrete. For singular continuous spectrum 
(as in SNAs \cite{qqq}) the scaling exponent satisfies $1<\beta<2$ 
\cite{qqq}. Here we find that the Fourier sum obeys scaling, with an 
exponent slightly greater than unity; see Figs.~\ref{fig6}((a) and (b)). This 
implies that the dynamics in these attractors is typically noisy unlike SNAs 
where the dynamical correlation persists over long times due to intermittency. 
For nonchaotic attractors in the H\'enon and the Lorenz systems, we find the 
exponents  $\beta=1.05$ and $\beta=1.04$ respectively. Fig.~\ref{fig6}((c) and 
(d)) shows the respective spectral trajectories in the complex plane (Re $X$,Im $X$).
\begin{figure*}
\scalebox{0.4}{\includegraphics{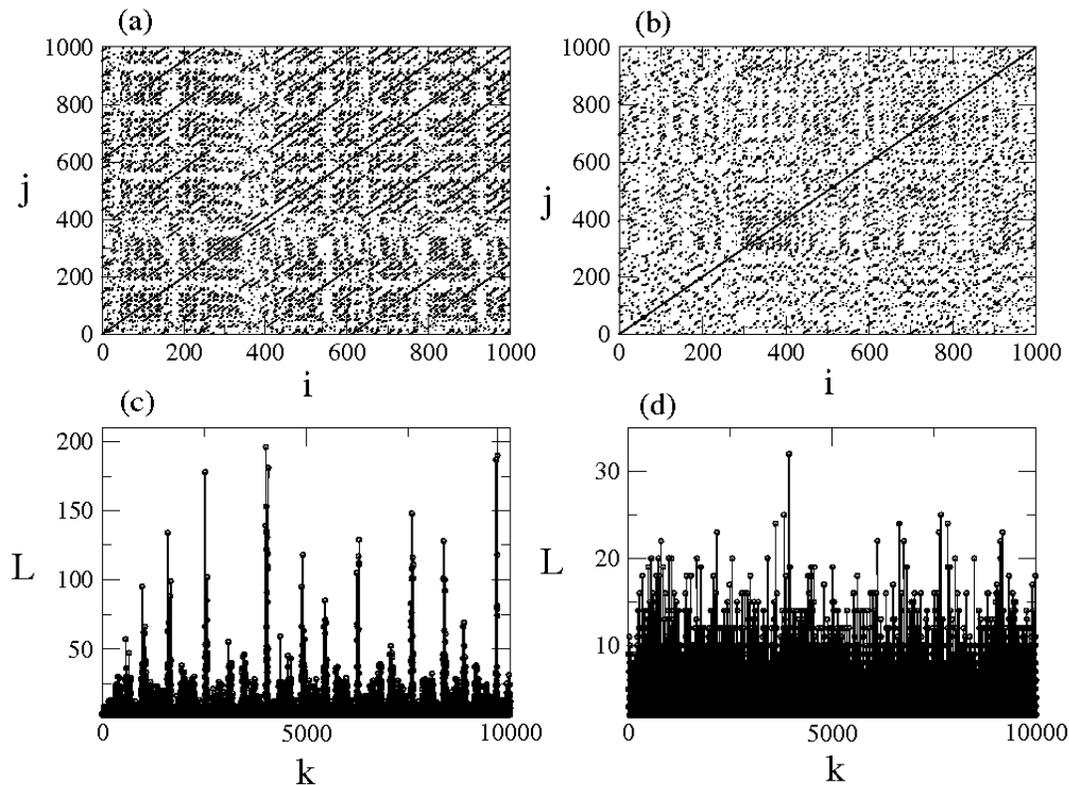}}
\caption{Comparision of recurrence plots for (a) a strange nonchaotic attractor 
for the quasiperiodically forced Henon map \cite{hen-sna}, and, (b) the aperiodic 
nonchaotic attractor for the driven H\'enon map with parameters as given in 
Fig.~\ref{fig1}(b). The RP for SNA has longer diagonal length segments due to long 
range correlations as compared to that of the ANA which consists of more isolated 
correlated points and shorter diagonal length segments. This is seen in (c) and (d) 
where the diagonal length distribution (with minimum length $L_{min}=2$ in each case) 
for the two cases are compared. We calculated the {\sl determinism} ($DET$) for 
both the cases (see Eq.~{46} in Marwan \etal\cite{rqa3}) and for SNA, $DET=0.78$, 
whereas for ANA it is $DET=0.62$. Here we take $\delta=0.3\sigma$ where $\sigma$ is 
the standard deviation of the trajectory.}
\label{fig7}
\end{figure*}

A detailed characterization of the nature of dynamics can be obtained from 
measures based on recurrences \cite{eckmann}. Recurrence plots (RPs) 
are defined for a given trajectory
$\{\vec{\mathbf{x}}_i\}_{i=1}^N$ through the matrix
\vspace{-0.2cm}
\begin{equation}\label{rp}
R_{i,j}=\Theta(\delta-||\vec{\mathbf{x}}_i - \vec{\mathbf{x}}_j||), 
\qquad i,j=1,\ldots,N
\end{equation}
where $\delta$ is a predefined threshold, $\Theta(\cdot)$ the Heaviside
function and $\parallel.\parallel $ the maximum norm. The maximum norm 
(also called infinity norm) of a vector $\vec{\mathbf{x}}$ of length $N$ is 
given by $\parallel \vec{\mathbf{x}} \parallel_{\infty} = max(|x_{1}|,\ldots,
|x_{N}|)$. Points that are closer (respectively further) than $\delta$ yield 
an entry ``1'' (respectively ``0'') in the matrix $R_{i,j}$. Then, the 
values ``1'' and ``0'' are depicted as black and white dot in a 
two--dimensional plot, providing a visual representation of the system 
dynamics. The RPs exhibit characteristic large scale and small scale patterns 
(called {\sl typology} and {\sl texture} respectively); these have been 
comprehensively reviewed recently \cite{rqa3}. The selection criteria for the 
threshold $\delta$ is discussed in details in a review by Marwan \etal 
\cite{rqa3}. Here we take $\delta$ in units of the standard deviation $\sigma$ 
of the trajectory.

In Fig.~\ref{fig7}, we compare the RPs for a SNA in the quasiperiodically 
forced Henon map \cite{hen-sna} and the ANA  for the modulated H\'enon map. 
It is clear that the RP of ANA (Fig.~\ref{fig7}(b),(d)) consists of more 
isolated correlated points and short diagonal segments depicting short-range 
correlations. On the other hand, the RP for SNA (Fig.~\ref{fig7}(a),(c)) 
has a larger distribution of longer diagonal line segments implying that 
correlation persists over long times -- a signature of quasiperiodic driving. 

\begin{figure*}
\scalebox{0.5}{\includegraphics{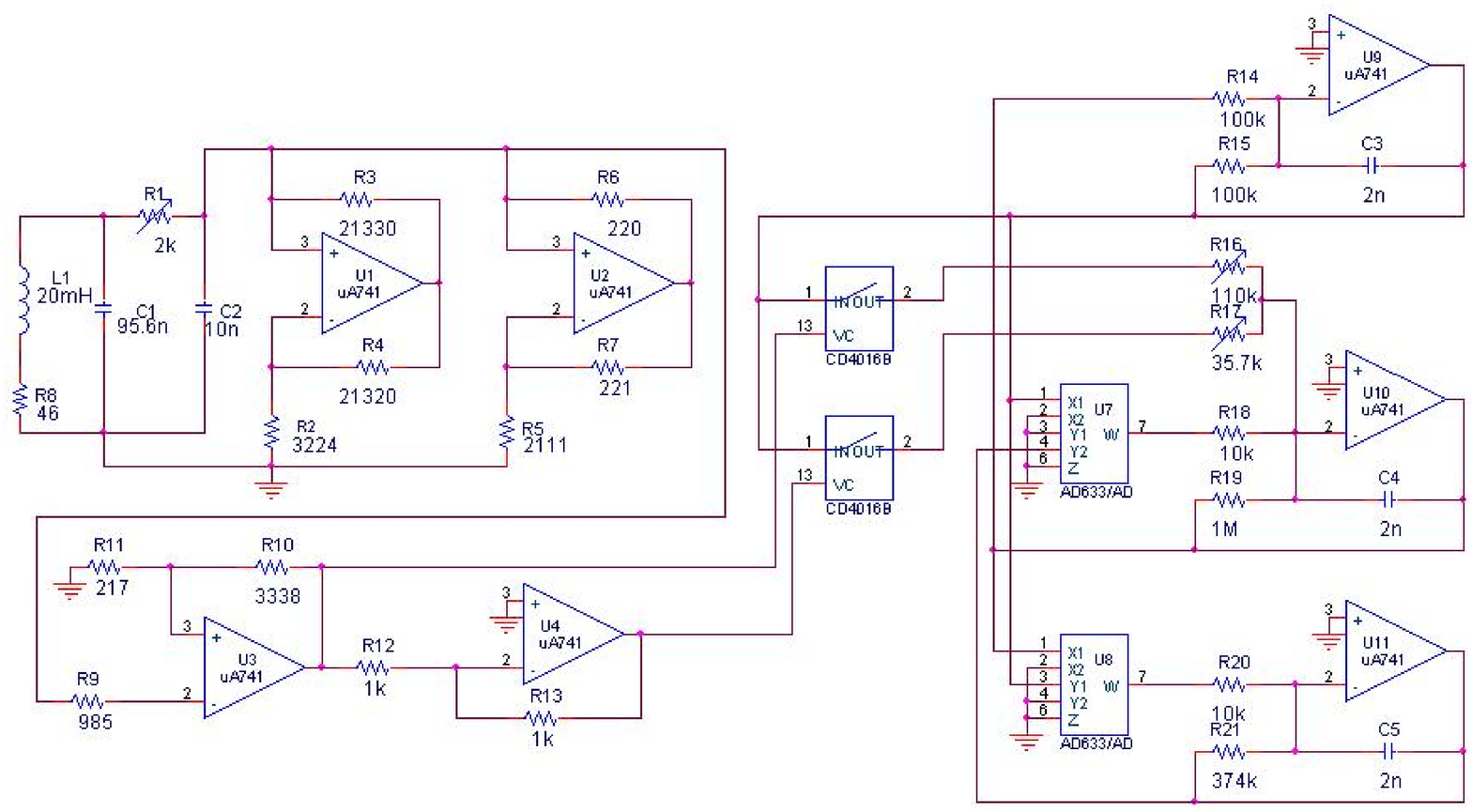}}
\caption{(Colour Online) 
Experimental Lorenz circuit: ICs (U1-U8) power supply $\pm9Volt$ and 
(U9-U11) power supply $\pm15$ Volt. All resistances are in Ohms.}
\label{fig8}
\end{figure*}

\section{Experiment}
\label{sec-exp}
In the electronic experiments reported here we have designed a circuit that 
essentially obeys the Lorenz equations \cite{Lorenz} and permits one of the 
parameters to be switched between two values. 
\begin{figure}
\scalebox{0.17}{\includegraphics{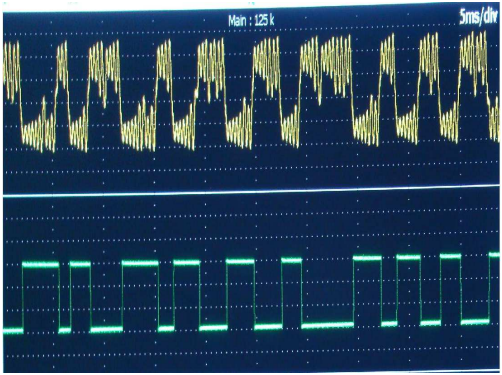}}
\caption{(Colour Online) Chaotic pulse generated using Chua oscillator: upper 
trace is the chaotic voltage VC$_2$ at capacitor node C2 of the Chua circuit 
and the  lower trace is the voltage measured at the output of the op-amp U3. 
Upper trace is scaled-up and the lower trace is scaled-down for visual 
clarity.}
\label{fig9}
\end{figure}

The typical parameters for the  butterfly attractor are  
$\sigma=10$, $b=8/3$ and in our circuit  the value 
of $\rho$ may be switched between 28 and 7 (as in Fig.~\ref{fig4}). 
The switching of $\rho$  is controlled by a chaotic pulse generated 
from a Chua circuit \cite{chua1,chua2}:  $\rho$ = 28  if the drive signal 
has value 1, else $\rho$= 7 (and the drive signal has the value 0). The 
experimental circuit of the pulse driven Lorenz oscillator is shown in 
Fig.~\ref{fig8}. A Chua circuit is designed using two op-amps 
(U1-U2: $\mu$741), capacitors C1 and C2, inductor L1 with a leakage 
resistance R8 and other resistances R1-R7. It generates a chaotic double 
scroll for choice components noted in the circuit diagram. The dynamics of 
the Chua circuit can be controlled by varying R1 resistance keeping other 
components fixed. The double-scroll chaos from the Chua circuit is then 
applied to a Schmitt trigger circuit designed by using op-amp U3, an inverting 
amplifier U4 and associated resistances R9-R13. The output from U3 and U4 
are used to control the analog switches U5A and U6A respectively to allow 
continuity of either R16 or R17 in the Lorenz circuit. The Lorenz circuit is 
implemented using two analog multipliers U7-U8, and three op-amps 
(U9-U11: $\mu$741), capacitors C3-C5 and resistances R14-R21. The choice of 
resistances R16 and R17 made the selection of $\rho$-value between 7 
($=\mbox{R19}/\mbox{R16}$) and 28 ($=\mbox{R19}/\mbox{R17}$) respectively. 
The other parameter of the Lorenz circuit are decided as 
$\sigma=\mbox{R19}/\mbox{R14}$ and $b=\mbox{R19}/\mbox{R21}$. The 
analog switches are in ON state if their control pulse at VC terminal is 
positive. So the analog switch U5A is in ON state when the output of U3 is 
positive but U6A is in OFF state.
\begin{figure}
\scalebox{0.3}{\includegraphics{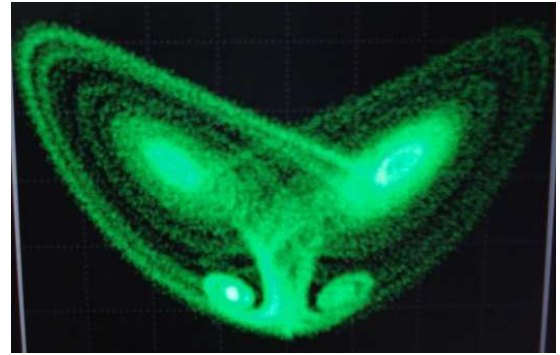}}
\caption{(Colour Online) 
Oscilloscope picture of the ANAs in the modulated Lorenz circuit: output 
voltage of U8 plotted against the output voltage of U11.}
\label{fig10}
\end{figure}

\begin{figure}
\scalebox{0.4}{\includegraphics{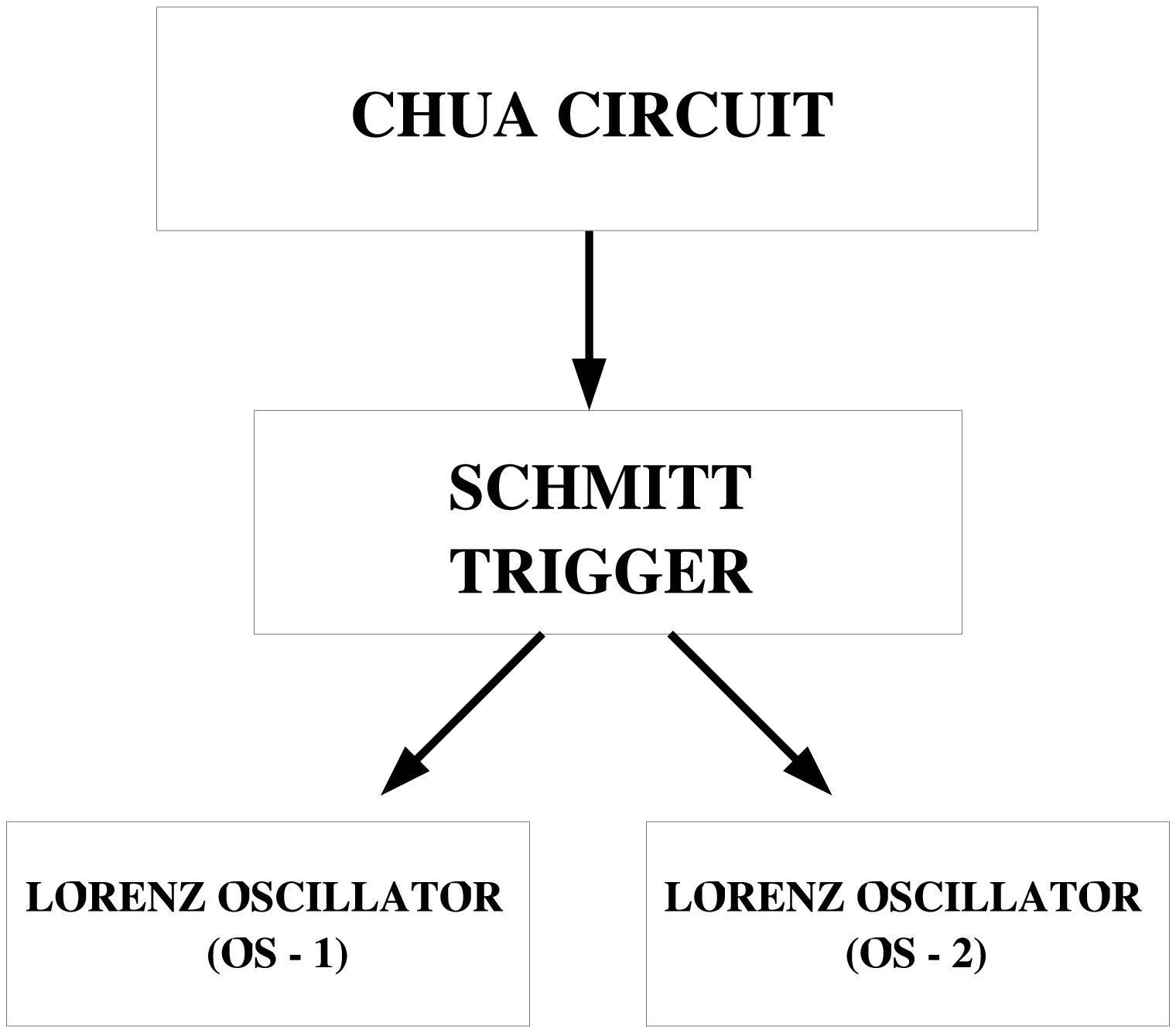}}
\caption{Block diagram of a system of two Lorenz circuits modulated by the 
chaotic drive of a Chua circuit.}
\label{fig11}
\end{figure}

Alternately, the analog switch U6A is in ON state and U5A is in OFF state 
when the output of U3 is negative but inverted by the U4 to make the
control pulse positive at the VC terminal of U6A. The oscilloscope picture 
of the control pulse as generated from the Chua circuit is shown 
in Fig.~\ref{fig9}. The upper trace is the double scroll chaotic signal 
from the Chua circuit which is processed by the Schmitt trigger U3. The 
chaotic pulse is clearly seen in the lower trace as switching between a 
positive and a negative value almost randomly; the signal is scaled 
down in the oscilloscope. 

The chaotic control signal switches the $\rho$-value of the Lorenz 
circuit aperiodically. The phase portrait of the Lorenz circuit is shown 
in Fig.~\ref{fig10}: this is the ANA that results from the theoretical 
strategy outlined in Sec~\ref{sec-cs}; see Fig.~\ref{fig4}.

\begin{figure}
\scalebox{0.27}{\includegraphics{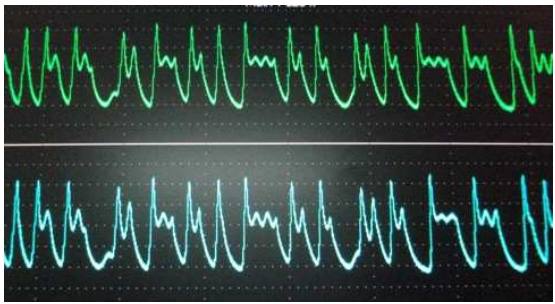}}
\caption{(Colour Online) Oscilloscope picture of the output voltages showing 
the experimental time series of the two Lorenz circuits.}
\label{fig12}
\end{figure}

\begin{figure}
\scalebox{0.29}{\includegraphics{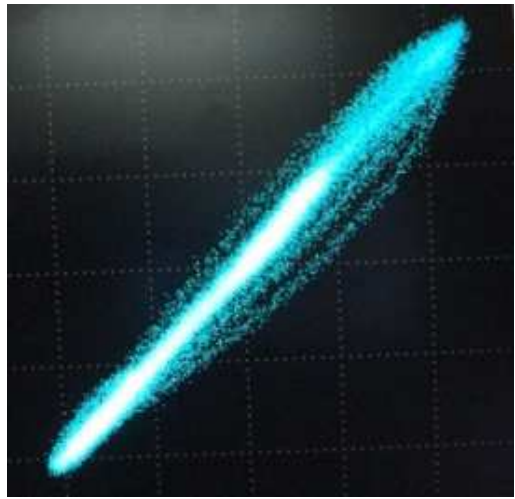}}
\caption{(Colour Online) 
Oscilloscope picture of synchronization: similar output signals as shown in 
Fig.~\ref{fig11} measured from two Lorenz circuits are plotted against each 
other to draw the synchronization manifold.}
\label{fig13}
\end{figure}
 
To show that the attractors obtained experimentally satisfy the 
synchronization condition, we constructed an auxiliary system 
\cite{gs}. Two identical Lorenz oscillators  are controlled by the chaotic 
pulse generated from a single Chua oscillator. The components of the Lorenz 
oscillators are carefully chosen with 1$\%$ tolerance so that both the 
oscillators are almost identical. Synchronization of two the nonchaotic 
Lorenz circuit is now investigated, the circuit scheme of which is shown in 
Fig.~\ref{fig11}. We observe that the output voltage of the Lorenz 
oscillators, OS-1 and OS-2, are completely synchronized. The oscilloscope 
pictures of the two time series from the oscillators are shown in green and 
blue in Fig.~\ref{fig12} for comparison. The time series plotted one 
against the other, is shown as a thick line, confirms complete synchronization 
of the two oscillators (within experimental bounds). The width of the 
synchronization manifold is due to natural parameter mismatch between the two 
designed Lorenz oscillators; this is unavoidable in experiments.

\section{Discussion}
\label{sec-dis}

Motion that is both stable and aperiodic is ubiquitous in natural systems 
\cite{pr1,glass}. The manner in which such dynamics can be created is 
therefore of interest. One class of attractors that have these features has 
been known for some time now, but a quasiperiodic drive is essential for 
their creation \cite{prasad} and thus these appear to be somewhat exceptional. 
An area where these considerations are potentially important is in the 
dynamics of biological systems. Although not manifestly periodic, several 
biological phenomena are stable, at least in a homeostatic sense 
\cite{glass,leibler}. Thus it is a moot question whether aperiodic but 
nonchaotic attractors are responsible for such stability.

On SNAs there is a delicate balance between global stability: as was 
established by Sturman and Stark \cite{sturman} there is an unstable set 
embedded within the attractor. The design strategy that we have enunciated 
in the present work keeps this feature in mind: the scheme we have proposed 
here is to modulate system parameters in such a manner as to achieve global 
stability while ensuring local instability.  

This method of dichotomous modulation creates attractors which are 
nonchaotic and have a fractal geometry on experimentally accessible 
timescales. We have recently shown \cite{snr} that SNAs created via quasiperiodic 
forcing are a manifestation of weak generalized synchronization, and that
similar stable attractors can be created by chaotic forcing \cite{snr}.
The parametric modulation used in the present case retains the
skew-product structure of the dynamical system. The formation of these stable 
attractors can in some sense be seen as an instance of generalized 
synchronization \cite{gs}. 

The attractors created via such parameter modulation are quite distinct 
from SNAs. The Gaussian nature of the FTLE distribution shows that the 
dynamics is not intermittent. From the spectral properties it is evident 
that unlike SNAs, the power spectrum varies as $|X(\Omega,N)|^{2}\sim N$, 
which occurs when the motion is random or chaotic. Our system being nonchaotic, 
the randomness in the motion comes from the stochastic nature of the 
modulation. This is confirmed by looking at the correlation properties via 
recurrence plots, which shows similiar behaviour to that of random or chaotic 
dynamics.

The aperiodic nonchaotic attractors can be realized in an experimental setup. 
As an example, we construct an electronic circuit experiment wherein a Chua double 
scroll attractor is used to drive a Lorenz attractor. The experiment closely 
matches the simulation result, and by experimentally constructing an auxiliary 
system \cite{aux} we demonstrate that for a given drive sequence, trajectories 
with different initial states synchronize rapidly on the ANAs. This demonstrates 
the possibility of creating---or using---such dynamics in practical applications 
\cite{secure}. Furthermore such a realization shows the robustness of the proposed 
design scheme against external noise. 

\begin{acknowledgments}
S.K.D. and S.K.B. acknowledges support by the DST, India under grant \#SR/S2/HEP-03/2005.
\end{acknowledgments}

\end{document}